\documentclass[10pt,twocolumn,twoside]{IEEEtran}

\usepackage{amsmath,amsfonts,amssymb,amscd,latexsym,enumerate,theorem,bbm}
\usepackage{arydshln}
\usepackage{graphicx,epsfig,psfrag}
\usepackage{subfigure}
\usepackage{tikz}
\usepackage{tkz-berge}
\usepackage{float}
\usepackage{cite}
\usepackage{dsfont}

\DeclareMathOperator{\im}{im}
\DeclareMathOperator{\col}{col}
\DeclareMathOperator{\diag}{diag}
\DeclareMathOperator{\bdiag}{blockdiag}

\let\emptyset\varnothing

\newcommand{\calE}{\ensuremath{\mathcal{E}}}

\newcommand{\calG}{\ensuremath{\mathcal{G}}}
\newcommand{\calH}{\ensuremath{\mathcal{H}}}
\newcommand{\calI}{\ensuremath{\mathcal{I}}}
\newcommand{\calJ}{\ensuremath{\mathcal{J}}}

\newcommand{\calR}{\ensuremath{\mathcal{R}}}

\newcommand{\calV}{\ensuremath{\mathcal{V}}}
\newcommand{\calW}{\ensuremath{\mathcal{W}}}

\newcommand{\hatp}{\ensuremath{\hat{p}}}

\newcommand{\hatB}{\ensuremath{\hat{B}}}

\newcommand{\bmat}{\begin{matrix}}
\newcommand{\emat}{\end{matrix}}
\newcommand{\bbm}{\begin{bmatrix}}
\newcommand{\ebm}{\end{bmatrix}}
\newcommand{\bpm}{\begin{pmatrix}}
\newcommand{\epm}{\end{pmatrix}}
\newcommand{\bse}{\begin{subequations}}
\newcommand{\ese}{\end{subequations}}
\newcommand{\beq}{\begin{equation}}
\newcommand{\eeq}{\end{equation}}
\newcommand{\ben}{\begin{enumerate}}
\newcommand{\een}{\end{enumerate}}
\newcommand{\beni}{\renewcommand{\labelenumi}{\roman{enumi}.}
\renewcommand{\theenumi}{\roman{enumi}}\begin{enumerate}}
\newcommand{\eeni}{\end{enumerate}\renewcommand{\labelenumi}{\arabic{enumi}.}
\renewcommand{\theenumi}{\arabic{enumi}}}
\newcommand{\bena}{\renewcommand{\labelenumi}{\alpha{enumi}.}
\renewcommand{\theenumi}{\alpha{enumi}}\begin{enumerate}}
\newcommand{\eena}{\end{enumerate}\renewcommand{\labelenumi}{\arabic{enumi}.}
\renewcommand{\theenumi}{\arabic{enumi}}}
\newcommand{\bit}{\begin{itemize}}
\newcommand{\eit}{\end{itemize}}
\newcommand{\bthe}{\begin{theorem}}
\newcommand{\ethe}{\end{theorem}}
\newcommand{\blem}{\begin{lemma}}
\newcommand{\elem}{\end{lemma}}
\newcommand{\bprop}{\begin{proposition}}
\newcommand{\eprop}{\end{proposition}}
\newcommand{\bex}{\begin{example}}
\newcommand{\eex}{\end{example}}
\newcommand{\bas}{\begin{assumption}}
\newcommand{\eas}{\end{assumption}}
\newcommand{\bre}{\begin{remark}}
\newcommand{\ere}{\end{remark}}
\newcommand{\bcor}{\begin{corollary}}
\newcommand{\ecor}{\end{corollary}}
\newcommand{\bdfn}{\begin{definition}}
\newcommand{\edfn}{\end{definition}}

\newcommand{\seq}[1]{\ensuremath{1,2,\ldots,#1}}

\newcommand{\R}{\ensuremath{\mathbb R}}

\newcommand{\BP}{\noindent{\bf Proof. }}
\newcommand{\EP}{\hspace*{\fill} $\blacksquare$\smallskip\noindent}

\newcommand{\EE}{\hspace*{\fill}$\square$}

\usepackage{multicol}
\usepackage{float}
\newcommand{\genset}[3]{
  \node[circle,draw,drop shadow,thick,minimum width=7mm,inner sep=0pt,fill=white,#3] (#1) at (#2) {};
  \draw[thick] ($(#2)-(2mm,0)$) sin ++(1mm,1mm) cos ++(1mm,-1mm) sin
  ++(1mm,-1mm) cos ++(1mm,1mm); }

\usepackage[prependcaption,colorinlistoftodos]{todonotes}

{\theorembodyfont{\itshape}\newtheorem{theorem}{Theorem}}
{\theorembodyfont{\itshape}\newtheorem{lemma}[theorem]{Lemma}}
{\theorembodyfont{\itshape}\newtheorem{proposition}[theorem]{Proposition }}
{\theorembodyfont{\itshape}\newtheorem{corollary}[theorem]{Corollary}}
{\theorembodyfont{\upshape}\newtheorem{definition}[theorem]{Definition}}
{\theorembodyfont{\itshape}}
{\theorembodyfont{\upshape}\newtheorem{remark}[theorem]{Remark}}
{\theorembodyfont{\upshape}}
{\theorembodyfont{\upshape}\newtheorem{example}[theorem]{Example}}
{\theorembodyfont{\upshape}\newtheorem{assumption}{Assumption}}

\newcommand{\ones}{\mathds{1}}

\newcommand{\part}[2]{\ensuremath{\frac{\partial #1}{\partial #2}}}

\newcommand{\fpart}[2]{\displaystyle \frac{\partial #1}{\partial #2}}

\usepackage{tikz}
\usepackage{tkz-berge}
\usetikzlibrary{positioning}
\usetikzlibrary{arrows,%
                petri,%
                topaths}%
\usetikzlibrary{calc,shadows}         
\tikzstyle{vertex}=[circle, shading = ball, ball color = white!100!white, minimum size = 15pt, draw, inner sep=0pt]  
\newcommand{\vertex}{\node[vertex]}           
\newcommand{\weight}[1]{{\footnotesize $\mathit{#1}$}}

\newcommand{\new}[1]{\textcolor{black}{#1}}

\title{\LARGE \bf
A Novel Reduced Model for Electrical Networks\\ with Constant Power Loads}

\author{Nima Monshizadeh\; \and \;Claudio De Persis \;\and\; Arjan J. van der Schaft \;\and\;  Jacquelien M.A. Scherpen%
\thanks{Nima Monshizadeh, Claudio De Persis, and Jacquelien M.A. Scherpen are with the Engineering and Technology Institute, University of Groningen, 9747AG, The Netherlands, {\tt\small n.monshizadeh@rug.nl, c.de.persis@rug.nl, j.m.a.scherpen@rug.nl}}
 \thanks{Arjan J. van der Schaft is with the Johann Bernoulli Institute for Mathematics and Computer Science, University of Groningen, 9700AK, The Netherlands, {\tt\small a.j.van.der.schaft@rug.nl}}%
 \thanks{This work is supported by NWO within the program ``Uncertainty Reduction in Smart Energy Systems (URSES)" under the auspices of the project ENBARK}}
\begin{document}
	\maketitle
\begin{abstract}
We consider a network-preserved model of power networks with proper algebraic constraints resulting from constant power loads.
Both for the linear and the nonlinear differential algebraic model of the network, we derive explicit reduced models which are fully expressed in terms of ordinary differential equations. For deriving these reduced models, we introduce the ``projected incidence" matrix which yields a novel decomposition of the reduced Laplacian matrix. With the help of this new matrix, we provide a complementary approach to Kron reduction which is able to cope with constant power loads and nonlinear power flow equations.
\end{abstract}

\section{Introduction}
The interdisciplinary field of power networks and microgrids has received lots of attention from the control community in the last decade, see e.g. \cite{dorfler2014breaking, JWSP-FD-FB:12u, schiffer2014conditions, trip2016internal, stegink2015unifying, Claudio-Nima-modular}.
Principal components of a power grid are synchronous generators, inverters, and loads. 
The frequency behavior of the synchronous generators is often modeled by the so called ``swing equation" \cite{Machowski2009}.
The frequency of the droop-controlled inverters also admits similar dynamics, see e.g. \cite{schiffer2014conditions}.

In a natural modeling of the power network, the generators and the loads are located at different subsets of nodes. This corresponds to the {\em structure  preserving} or {\em network preserving} model which is naturally expressed in terms of differential algebraic equations (DAE), see \cite{bergen1981structure}, \cite{dib2009globally}. The algebraic constraints in these models represent the load {characteristics}.

The methods that have been suggested to study and analyze these network-preserved models can be classified into two distinct categories. The first one is to directly use the differential algebraic model of the power network. This is typically done by studying the local solvability of the load power flow by the implicit function theorem and looking into the associated load flow Jacobian \cite{bergen1981structure,JS-FD:15,DP-Nima-Jo-Flo-CDC16}. 
The second approach, which will be pursued in this manuscript, relies on the derivation of a  {\em network reduced} model. 
Along this line of research, several aggregated models are reported in the literature where each bus of the grid is associated with certain load and generation; see e.g. \cite{chakrabortty2011measurement}, \cite{ourari2006dynamic}. The main advantage of these aggregated models is that they are described by ordinary differential equations (ODE) which facilitates the analysis and numerical simulation of the network. However, the explicit relationship between the aggregated model and the original structure preserved model is often missing, which restricts the validity and applicability of the results. Aiming at a simplified ODE description of the model together with respecting the heterogeneous structure of the power network  has popularized the use of Kron reduced models \cite{kron:39,dorfler2013kron,caliskan2014towards}. In the Kron reduction method, the variables which are exclusive to the algebraic constraints are solved in terms of the rest of the variables. This results in  a reduced graph whose (loopy) Laplacian matrix is the Schur complement of the (loopy) Laplacian matrix of the original graph. Notice that unlike the aggregated models, the Kron reduced ones are obtained by explicitly solving the algebraic equations associated with the loads. Despite the attractive simplicity and the interesting theory, the Kron reduction modeling essentially restricts the class of the applicable load models to constant admittance loads and current sources  \cite{dorfler2013kron,dorfler2013synchronization}.

Algebraic constraints present in the differential algebraic model of the power network can also be solved in the case of frequency dependent loads where the active power drawn by each load consists of a constant term and a frequency-dependent term \cite{bergen1981structure}, \cite{zhao2015distributed}, \cite{monshizadeh2015output}.  However, in the popular class of constant power loads, the algebraic constraints are ``proper", meaning that they are not explicitly solvable. To the best of our knowledge, for nonlinear power networks with proper algebraic constraints, an {\em explicit} reduced ODE model is absent in the literature.\footnote{with the exception of the abridged version of this work 
\cite{nima-reduced-acc16}.}

In this paper, first we revisit the Kron reduction method for the linear case, where the Schur complement of the Laplacian matrix (which is again a Laplacian) naturally appears in the network dynamics. It turns out that the usual decomposition of the reduced Laplacian matrix leads to a state space realization which contains merely partial information of the original power network, and the frequency behavior of the loads is not immediately visible. Moreover, this decomposition does not provide useful insight for the nonlinear model which is the main focus of the current manuscript. As a remedy for this problem, we introduce a new matrix, namely the projected incidence matrix, which yields a novel decomposition of the reduced Laplacian. Then, we derive reduced models capturing the behavior of the original network-preserved model. Next, we turn our attention to the nonlinear case where the algebraic constraints are not readily solvable. Again by the use of the projected incidence matrix, we derive explicit reduced models expressed in terms of ordinary differential equations. We identify the loads embedded in the derived reduced network by unveiling a conserved quantity of the system. Furthermore,  we carry out the Lyapunov stability analysis of the proposed reduced model together with a distributed averaging controller guaranteeing frequency regulation and power sharing.

The structure of the paper is as follows. Section \ref{s:powernet} describes the power network model we consider in this paper. In Section \ref{s:lin}, we discuss the reduced models for the system obtained by linear approximations. In addition, we introduce the projected incidence matrix and the new decomposition of the reduced Laplacian matrix. An explicit reduced model for the nonlinear power network is established in Section \ref{s:nl}. Stability of the model together with power sharing is discussed in Section \ref{s:convergence}. Finally, the paper closes with conclusions in Section \ref{s:concl}.  

\section{Power Network}\label{s:powernet}
The topology of the power network is represented by a connected and undirected graph $\mathcal{G}(\mathcal{V},\mathcal{E} )$. There are two types of buses (nodes): generators $\calV_G $ and loads $\calV_L$, with $\mathcal{V}=\mathcal{V}_G \cup \mathcal{V}_L$. The number of generators and loads are denoted by $n_g$ and $n_\ell$, respectively. The edge set $\mathcal{E}$ is the set of unordered pairs $\{i,j\}$ accounting for the transmission lines which are assumed to be inductive. The total number of nodes is denoted by $n$, and that of the edges by $m$.
Let the matrix $B$ denote the incidence matrix of $\calG$. Recall that for an undirected graph $\calG$, the incidence matrix is obtained by assigning an arbitrary orientation to the edges of $\calG$ and defining 
\[ b_{ik} = \left\{
  \begin{array}{l l}
    +1 & \quad \text{if $i$ is the tail of arc $k$}\\
    -1 & \quad \text{if $i$ is the head of arc $k$}\\
    0 & \quad \text{otherwise}
  \end{array} \right.\]
 with $b_{ik}$ being the $(i,k)^{th}$ element of $B$.

 At each node $i\in \calV$, the electrical active power  is given by
\beq\label{e:active-i}
p_i = \sum_{j\in \mathcal{N}_i} X_{ij}^{-1} V_iV_j \sin \theta_{ij},\quad \theta_{ij}:= \theta_i-\theta_j
\eeq
where $X_{ij}$ is the inductance of the transmission line $\{i,j\}$, $V_i$ is the voltage magnitude at node $i$, and $\theta_i$ is the voltage angle with respect to the nominal reference $\theta^\ast=\omega^\ast t$. We assume that the transmission lines are lossless and the voltage magnitudes are constant. 
We consider generators admitting the so-called swing equation 
\beq\label{e:gen-i} 
M_i\ddot{\theta}_i=-A_i\dot\theta_i-p_i+u_i, \qquad i\in \calV_G
\eeq
where $M_i$ is the angular momentum, $A_i$ is the damping coefficient, and $u_i$  is the controllable power generation. 
%
The dynamics \eqref{e:gen-i} can both model synchronous generators \cite{machowski2011power} and droop-controlled inverters with virtual inertia \cite{soni2013improvement, bevrani2014virtual} or power measurement delays \cite{schiffer2014conditions}. In the case of inverters, $M_i$ is the power measurement delay or virtual inertia, and $A_i$ is a tunable {droop control} gain. {For simplicity, in what follows we use the term ``generator'' for either case.}

As for the loads, we consider the constant active power loads admitting the algebraic constraint 
\[
0=p_i-p_i^\ast, \qquad i\in \calV_L
\]
where $p_i^\ast$ is constant. 
Now the network model can be written in compact form as
\bse\label{e:nonlinear}
\begin{align}
\label{e:gen}
M\ddot{\theta}_G&=-A\dot\theta_G-B_G\Gamma{\rm sin}(B^T\theta)+u\\
\label{e:load}
0&=-B_L\Gamma{\rm sin}(B^T\theta)+p^\ast
\end{align}
\ese
where $\theta_G=\col(\theta_i)$ with $i\in\calV_G$, and $\theta_L=\col(\theta_i)$ with $i\in \calV_L$.
The $\sin(\cdot)$ operator is interpreted element-wise. 
In addition, $\theta=\col(\theta_G, \theta_L)$, $B=\col(B_G, B_L)$, and $\Gamma=\diag(\gamma_k)$ with 
\[
\gamma_k=X_{ij}^{-1}V_iV_j,
\]
where $k$ is the index of the edge $\{i, j\}$ in accordance with the incidence matrix 
$B$. The notation $\col(Y_1, Y_2)$ is used to denote in short the matrix $\bbm Y_1^T& Y_2^T\ebm^T$ for given matrices $Y_1$ and $Y_2$. 
Again note that the voltages are assumed to be positive and constant, and thus the matrix $\Gamma$ is positive definite.  
This is consistent with the standard decoupling assumption \cite{JWSP-FD-FB:12u,dorfler2014breaking,monshizadeh2015output,JS-FD:15}.
Our goal in this paper is to eliminate the {load flow equations} and embed them into the dynamics of the generators in order to obtain an explicit reduced model described by ordinary differential equations. 

\section{Linear model}\label{s:lin}
First, we consider the linear model where $\sin(\eta)$ is approximated by $\eta$, with $\eta=B^T\theta$. This means that the differences of the phase angles are assumed to be relatively small, which is satisfied in a vicinity of the nominal condition. Then, the system \eqref{e:gen}-\eqref{e:load} can be written as
\medskip{}
\begin{align}\label{e:system-lin}
\bbm
M\ddot{\theta}_G+A\dot\theta_G\\[1mm]
0
\ebm
=-
\bbm
B_G\Gamma B_G^T &B_G\Gamma B_L^T\\[1mm]
B_L\Gamma B_G^T &B_L\Gamma B_L^T
\ebm
\bbm
\theta_G\\[1mm]
\theta_L
\ebm
+
\bbm
u\\[1mm]
p^\ast
\ebm
\end{align}
Note that the two by two block matrix on the right-hand side of \eqref{e:system-lin} can be associated with the Laplacian matrix of the graph $\calG$ with weights $\Gamma$: 
\[
L=B\Gamma B^T.  
\]
By \eqref{e:system-lin}, the vector $\theta_L$ can be computed as
\beq\label{e:thetaL}
\theta_L=-(B_L\Gamma B_L^T)^{-1}B_L\Gamma B_G^T\theta_G+(B_L\Gamma B_L^T)^{-1}p^\ast.
\eeq
Note that $B_L\Gamma B_L^T$ is a principal submatrix of the Laplacian matrix and thus invertible. By replacing this back to \eqref{e:system-lin}, we obtain 
\beq\label{e:red-lin}
M\ddot{\theta}_G=-A\dot\theta_G-L_S\theta_G+u-\hat{p}
\eeq
where 
$$
L_S=B_G\Gamma B_G^T-B_G\Gamma B_L^T(B_L\Gamma B_L^T)^{-1}B_L\Gamma B_G^T
$$
and 
\beq\label{e:phat}
\hat{p}=B_G\Gamma B_L^T(B_L\Gamma B_L^T)^{-1}p^\ast.
\eeq
The matrix $L_S$ is equal to the Schur complement of the Laplacian matrix $L$. It is well-known (see \cite[Lem. 2.1]{dorfler2013kron}) that $L_S$ is again a Laplacian matrix defined on a reduced graph $\hat{\cal{G}}=(\calV_G, \hat{\calE})$, and admits the decomposition 
\beq\label{e:cust}
L_S=\hatB \hat{\Gamma} \hatB^T
\eeq
where $\hatB$ is the incidence matrix of $\hat\calG$. 

A crucial issue in frequency regulation is to keep the frequency disagreement among the buses as small as possible, and steer the frequency back to the nominal frequency using a secondary control scheme. Notice that this frequency disagreement is not transparent in \eqref{e:red-lin}. Now, let $\omega_G=\dot{\theta}_G$, $\omega_L=\dot{\theta}_L$, and $\omega=\col(\omega_G, \omega_L)$. 
To capture the frequency disagreements in the original network \eqref{e:gen}-\eqref{e:load}, we define the vector ${v}\in \R^m$ as
\beq\label{e:delta}
v=B^T\omega.
\eeq
Observe that ${v}_k$ indicates the difference between the (actual) frequencies of the nodes $i$ and $j$, with $\{i, j\}$ being the {$k-\text{th}$} edge of $\calG$. Similarly, let the vector $\eta\in \R^m$ be defined as $\eta=B^T\theta.$
Then the network dynamics \eqref{e:gen}-\eqref{e:load} admits the following linear differential algebraic model 
\bse\label{e:sys-ss}
\begin{align}
\label{e:sys-eta}
\dot{\eta}&={v}=B^T\omega\\
M\dot{\omega}_G&=-A\omega_G-B_G \Gamma \eta+u\\
0&=-B_L\Gamma \eta+p^\ast.
\end{align}
\ese
{\textcolor{black}{We remark that the vector $\theta$ is defined on the nodes, whereas $\eta$ components live in the edge space.
Exploiting the latter variables in representing physical systems defined on graphs is ubiquitous in passivity and port-Hamiltonian based modeling; see e.g. \cite{burger.depersis.aut15,Arcak2007,van2014port,schaft2013,monshizadeh2015output,van2016perspectives} and Remark \ref{r:ph}.}}

Similarly, for the ODE model \eqref{e:red-lin}, we define the frequency disagreement vector as 
\beq\label{e:bad}
\hat{v}=\hatB^T\omega_G.
\eeq
Then by \eqref{e:cust} the system \eqref{e:red-lin} has the following state-space representation
\bse\label{e:agg}
\begin{align}
\label{e:agg-eta}
\dot{\hat\eta}&=\hat{v}=\hatB^T\omega_G\\
M\dot{\omega}_G&=-A\omega_G-\hatB \hat\Gamma  \hat{\eta}+u-\hatp
\end{align}
\ese
where $\hat\eta=\hatB^T\theta_G$. 

Although the Kron reduced model \eqref{e:agg} provides an explicit reduced model for the network \eqref{e:sys-ss}, comparing the dynamics \eqref{e:agg-eta} to \eqref{e:sys-eta} reveals certain disadvantages for this model.
First, unlike the vector ${v}$, the disagreement vector $\hat{v}$ captures only the mismatch among the frequencies of the generators, whereas, clearly one would like to monitor the mismatch of the frequencies in the entire network. Furthermore, the graph $\hat\calG$ is in general a complete graph, and hence the vectors $\hat\eta$ and $\hat{v}$ have $\frac{n_g^2-n_g}{2}$ elements. Therefore, the size of these vectors increases substantially by the increase in the size of the network, which makes the monitoring and simulations intractable. Finally, and most importantly for this work, the representation \eqref{e:agg} does not extend to the nonlinear model \eqref{e:nonlinear}. 


Motivated by the above drawbacks, next we propose an alternative decomposition of the reduced Laplacian matrix $L_S$, instead of the customary one given by \eqref{e:cust}.

\subsection{A novel decomposition of the reduced Laplacian}\label{ss:novel}
We make the result of this subsection self contained, and independent of the power network interpretation. 
To this end, let again $\calG=(\calV, \calE)$ denote an undirected graph with $n$ vertices and $m$ edges, and assume that $\calG$ is connected. As before, for each $k=\seq{m}$, let $\gamma_k>0$ denote the weight associated to the $k^{th}$ edge of $\calG$.
The Laplacian matrix of $\calG$ is defined as 
$
L=B\Gamma B^T
$
where $B$ is the incidence matrix and $\Gamma=\diag(\gamma_k)$. We refer to the matrix $\Gamma$ as the weight matrix, and we allow it in general to be time-dependent as long as the weights remain positive for all time.  Suppose that the vertex set $V$ is partitioned as $\calV=\calV_1 \cup \calV_2$ with $\calV_1 \cap \calV_2=\emptyset$. Then the Laplacian matrix $L$ can be partitioned as
$$
L=\bbm
L_{11} & L_{12}\\[1mm]
L_{12}^T& L_{22}
\ebm
$$
where $L_{11}\in \R^{|\calV_1|\times |\calV_1|}$.
Note that the Schur complement of $L$ with respect to $L_{22}$ is given by
$$
L_S=L_{11}-L_{12}L_{22}^{-1}L_{12}^T.
$$
This can be rewritten as
\beq\label{e:LS}
L_S=B_1\Gamma B_1^T-B_1\Gamma B_2^T(B_2\Gamma B_2^T)^{-1}B_{2}\Gamma B_1^T
\eeq
where $B=\col(B_1, B_2)$ is partitioned in accordance with the partitioning of $\calV$.
Again note that, for a connected graph $\calG$, the matrix $L_S$ is well defined, and is the Laplacian matrix of an undirected graph $\hat{\calG}$ with $|\calV_1|$ vertices. Now we have the following key definition:
\begin{definition}\label{d:Bs}
With respect to a partitioning $B=\col(B_1, B_2)$ and a weight matrix $\Gamma$, we define the {\em projected incidence matrix} {$B_S\in \R^{|\calV_1|\times m}$} as 
\beq\label{e:schur-inc}
B_S=B_1(I-B_2^+B_2)
\eeq
where $B^+_2$ is a {\color{black}{right inverse{\footnote{this is well-defined as $\calG$ is connected, and thus $B_2$ has full row rank.}} of the matrix $B_2$ given by}} 
\[
B^+_2=\Gamma B_2^T (B_2 \Gamma B_2^T)^{-1}.
\]
\end{definition}\EE

Observe that 
$B_S=B_1\Pi$
where 
\beq\label{e:Pi}
\Pi=I-B^+_2 B_2
\eeq
is the orthogonal projection to the kernel of $B_2$, with respect to the inner product defined by $\Gamma$.  
Each column of the matrix $B_S$ may have more than two nonzero elements, however, it has zero row sums similar to the incidence matrix. Some useful properties of the projected incidence matrix are captured in the following proposition:
\begin{proposition}\label{p:Bs}
{As in \eqref{e:schur-inc}}, let $B_S$ denote the projected incidence matrix of $\calG$ with respect to the partitioning $B=\col(B_1, B_2)$ and the weight matrix $\Gamma$. Then the following statements hold:
\medskip{}
\begin{enumerate}[(i)]
\item $\im \ones=\ker B_S^T$ (zero row sums)
\medskip{}
\item $0=B_S\Gamma B_2^T$
\medskip{}
\item $L_S=B_S\Gamma B_1^T$
\medskip{}
\item $L_S=B_S\Gamma B_S^T$ (new decomposition)
\medskip{}
\end{enumerate}
\end{proposition}
\BP
Clearly,
\begin{align}\label{e:2nd-s}
\nonumber
B_S\Gamma B_2^T&=B_1(I-B^+_2B_2)\Gamma B_2^T\\
&=B_1\Gamma B_2^T-B_1B^+_2B_2\Gamma B_2^T=0,
\end{align}
which proves the second statement.
From \eqref{e:LS}, we have
\begin{align*}
L_S&=B_1(I-\Gamma B_2^T(B_2 \Gamma B_2^T)^{-1}B_2)\Gamma B_1^T\\
&=B_1(I-B^+_2B_2)\Gamma B_1^T=B_S\Gamma B_1^T,
\end{align*}
which verifies the third statement. 


The matrix $B_S\Gamma B_S^T$ is computed as
\beq\label{4th-s}
B_S\Gamma B_S^T=B_S\Gamma B_1^T-B_S\Gamma B_2 ^T (B_2^+)^T B_1^T.
\eeq
By the third statement of the proposition, $B_S\Gamma B_1^T=L_S$. 
In addition, the second term on the right-hand side of \eqref{4th-s} is equal to zero by \eqref{e:2nd-s}. 
Therefore, we obtain that $B_S\Gamma B_S^T=L_S$.

As the matrix $L_S$ is the Laplacian matrix of a reduced graph $\hat{\calG}$, we have $L_S\ones=0$. Then, by the forth statement of the proposition and positive definiteness of $\Gamma$, we have $B_S^T\ones =0$. Recall that $L_S$ is the Schur complement of the Laplacian matrix $L$. As $\calG$ is connected, the spectral interlacing property \cite[Thm. 3.1]{fan2002schur} implies that $\hat{\calG}$ is connected as well, and thus $\ker L_S=\ker B_S^T=\im \ones$.
\EP
\begin{figure}
\[\begin{tikzpicture}[x=1.7cm, y=1.2cm,
    every edge/.style={sloped, draw, line width=1.2pt}]

\vertex (v1) at (-1,1)  {\small $1$};
\vertex (v2) at (1,1) {\small $2$};
\vertex [ball color=blue!20!white](v3) at (0,0) {\small $3$};
%
\path
(v1) edge node[anchor=north]{\weight{a}}(v3)
(v2) edge node[anchor=north]{\weight{b}}(v3);
\end{tikzpicture}
\]
\caption{Graph $\calG$ in Example \ref{ex}.}\label{f:graph}
\end{figure}

\begin{example}\label{ex}
Consider the graph $\mathcal{G}$ with three nodes in Figure \ref{f:graph}. The vertex set is given by $V=\{1, 2, 3\}$ which is partitioned as $V_1=\{1, 2\}$, 
$V_2=\{3\}$.  The weight matrix $\Gamma$ is equal to $\diag(a, b)$, with $a, b>0$. An incidence matrix of $\mathcal{G}$ is obtained as
\[
B=
\left[
    \begin{array}{cc}
    1&0\\
    0& -1\\ \hdashline[2pt/2pt]
    -1 &1
    \end{array}
\right]=
\bbm B_1 \\ B_2 \ebm.
\]
Now the projected incidence matrix is computed as
\[
B_S=
\left[
    \begin{array}{cc}
\dfrac{b}{a+b}&\dfrac{a}{a+b}\\[3.5mm] \dfrac{-b}{a+b}&\dfrac{-a}{a+b}
    \end{array}
\right].
\]
Clearly, the matrix above has zero row sums. 
Besides, the Laplacian matrix of $\calG$, and its Schur complement are obtained as
\[
L= \left[
    \begin{array}{cc;{2pt/2pt}c}
    a&0&-a\\
    0&b&-b\\ \hdashline[2pt/2pt]
    -a&-b&a+b
    \end{array}
\right],
\qquad L_S=
\left[
    \begin{array}{cc}
\dfrac{ab}{a+b}&\dfrac{-ab}{a+b}\\[3.5mm] \dfrac{-ab}{a+b}&\dfrac{ab}{a+b}
    \end{array}
\right],
\]
respectively. In agreement with Proposition \ref{p:Bs}, it is easy to verify that both $B_S\Gamma B_1^T$ and $B_S\Gamma B_S^T$ yield the same expression as $L_S$ given above.\EE
\end{example}

\subsection{A new representation of the reduced network}
Consider again the model \eqref{e:red-lin}. Let $B_S$ be the projected incidence matrix with respect to the partitioning $B=\col(B_G, B_L)$ and the weight matrix $\Gamma$, as given by Definition \ref{d:Bs}. Note that the indices $1$ and $2$ in the algebraic results of the previous subsection need to be replaced here by $G$ and $L$, respectively, i.e., 
\beq\label{e:BS-again}
B_S=B_G(I-B_L^+ B_L), \quad B_L^+=\Gamma (B_L\Gamma B_L^T)^{-1}.
\eeq
Now, let the vector $\eta_S$ be defined as
\beq\label{e:good}
\eta_S=B_S^T\theta_G.
\eeq
Note that $\eta_S$ has the same size as $\eta$ in the DAE \eqref{e:sys-ss}, \new{since $B_S$ has the same number of columns as $B$.}
By \new{taking the time derivate of both sides of \eqref{e:thetaL} along any solution to \eqref{e:system-lin},} we have
\beq\label{e:wL-wG}
\omega_L=-(B_GB_L^+)^T\omega_G
\eeq
where again $B_L^+$ denote a right inverse of $B_L$ given by $B^+_L=\Gamma B_L^T (B_L \Gamma B_L^T)^{-1}$.
Now, we write the following important equality
\begin{align}\label{e:nima-important}
B_S^T\omega_G&=B^T\omega 
\end{align}
where we have used 
\begin{align*}
&B_S^T\omega_G=(I-B^+_LB_L)^TB_G^T\omega_G
\\&\quad=B_G^T\omega_G-B_L^T(B_GB^+_L)^T\omega_G
=B^T_G\omega_G+B^T_L\omega_L
\end{align*}
along with \eqref{e:wL-wG}.
Also note that, by Proposition  \ref{p:Bs}(iv), we have
\[
L_S\theta_G=B_S\Gamma B_S^T \theta_G=B_S\Gamma \eta_S.
\]
Therefore the system \eqref{e:red-lin} admits the following state space model
\bse\label{e:agg-new}
\begin{align}
\label{e:agg-new-eta}
\dot{\eta}_S&={v}=B_S^T\omega_G\\
M\dot{\omega}_G&=-A\omega_G-B_S \Gamma \eta_S+u-\hatp
\end{align}
\ese
where $\hatp$ and $v$ are given by \eqref{e:phat} and \eqref{e:delta}, respectively.
\new{
The relationship among the models \eqref{e:system-lin}, \eqref{e:red-lin}, \eqref{e:sys-ss}, \eqref{e:agg}, and \eqref{e:agg-new} is partially summarized in the following proposition:
\begin{proposition}
For given $u$ and $p^\ast$, let $\theta=\col(\theta_G, \theta_L)$ satisfy the differential algebraic equation \eqref{e:system-lin}. Then the following statements hold:
\begin{enumerate}[(i)]
\item $\theta_G$ is a solution to differential equation \eqref{e:red-lin}.
\medskip{}
\item $(\eta, \omega_G, \omega_L)$ with $\eta=B^T\theta$, $\omega_G=\dot\theta_G$, and $\omega_L=\dot{\theta}_L$ is a solution to  the differential algebraic equation \eqref{e:sys-ss}.
\medskip{}
\item $(\hat\eta, \omega_G)$ with $\hat\eta=\hat{B}^T\theta$ and $\omega_G=\dot\theta_G$ is a solution to \eqref{e:agg}.
\medskip{}
\item $(\eta_S, \omega_G)$ with $\eta_S=B_S^T\theta$ and $\omega_G=\dot\theta_G$ is a solution to \eqref{e:agg-new}.
\end{enumerate}
\end{proposition}}
\BP
\new{The proof follows by construction of the models discussed prior to the proposition.}
\EP

\new{Note that, similarly, one can start with solutions of the reduced ODE models and construct solutions for  the DAE model \eqref{e:system-lin} upon certain compatibility of initial conditions. To avoid repetition, we provide such converse relations only for the case of the nonlinear model \eqref{e:nonlinear}; see Theorem \ref{t:relation}.
}
\begin{remark}\label{r:ph}
The proposed reduced model \eqref{e:agg-new} admits the following port-Hamiltonian form (see e.g. \cite{van2014port}, \cite{schaft2013} for a more general perspective):  
\begin{align}\label{e:ph-structure}
\bbm
{\dot\eta_S}\\[1mm]
\dot{\rho}
\ebm
=
\underbrace{
\bbm
0&B_S^T\\[1mm]
-B_S&-A
\ebm}_{\calJ-\calR}
\bbm
\fpart{\calH}{\eta_S}\\[4mm]
\fpart{\calH}{\rho}
\ebm
+\bbm
0\\[1mm]
I
\ebm
(u-\hatp)
\end{align}
where $\rho=M\omega_G$ and the Hamiltonian is given by
\[
\calH(\eta_S, \rho)=\frac{1}{2}\rho^TM^{-1}\rho-\ones ^T \Gamma \cos(\eta_S).
\]
The reduced system \eqref{e:agg}, obtained from the usual decomposition of the Laplacian,  also admits a port-Hamiltonian representation similar to the above where $\eta_S$ and $B_S$ are replaced by $\hat\eta$ and $\hatB$, respectively.
The corresponding Hamiltonian in the latter case is given by
\[
\hat{\calH}(\hat\eta, \rho)=\frac{1}{2}\rho^TM^{-1}\rho-\ones ^T \hat\Gamma \cos(\hat\eta).
\]
Notice that the first term on the right-hand side of the above equality represents the Kinetic energy and appears both in $\calH$ and $\hat\calH$. On the other hand, while the second term of $\hat\calH$ is primarily algebraic, the one of $\calH$ is associated with the potential energy defined on the edge variables $\eta_S\in \R^m$, with the same weights, $\Gamma$, and the same edge space, $\R^m$, as the original graph $\calG$ of the network. 
\EE
\end{remark}

The main advantage of the reduced model \eqref{e:agg-new} over \eqref{e:agg} is that the model \eqref{e:agg-new} readily reflects the properties of the full network \eqref{e:sys-ss}. Notice that both the frequency disagreement vector ${v}$ and the weight matrix $\Gamma$ of the DAE \eqref{e:sys-ss} are identically preserved in the reduced model; see also the remark above. By \eqref{e:nima-important}, the subdynamics \eqref{e:agg-new-eta} readily demonstrates the time evolution of frequency in the entire network (including both generators and loads). Hence, one can easily deduce the behavior of the full network by looking into the model \eqref{e:agg-new}, and the aforementioned shortcoming of the model \eqref{e:agg} do not apply. In particular, the ability of the method to deal with the nonlinear system is dealt with in the next section.

\section{Nonlinear model}\label{s:nl}

In this section, we consider the nonlinear model \eqref{e:gen}-\eqref{e:load}, and investigate possible elimination of purely algebraic constraints resulting from the constant power loads \eqref{e:load}. Notice that unlike the linear case, the state components $\theta_L$ cannot be explicitly solved in terms of $\theta_G$ and $p$.

 Before proceeding with the derivation of a reduced model, it is necessary to assume that \eqref{e:gen}-\eqref{e:load} admits a solution. To make this assumption more explicit, we write the differential algebraic system \eqref{e:gen}-\eqref{e:load} as
 \bse\label{e:sys-nl-ss}
 \begin{align}
 \label{e:sys-nl-ss-thetaG}
 \dot\theta_G&=\omega_G\\
 \label{e:sys-nl-ss-omegaG}
 M\dot\omega_G&=-A\omega_G-B_G\Gamma \sin (B^T\theta)+u\\
\label{e:alg}
0&=-B_L\Gamma \sin (B^T\theta)+p^\ast.
 \end{align}
 \ese
Suppose that $\overline \theta=\col(\overline \theta_G. \overline \theta_L)$  and $\overline u$ are constant vectors satisfying
\bse\label{e:ss-conditions}
\begin{align}
0&=-B_G\Gamma \sin (B^T \overline \theta)+\overline u\\
 \label{e:alg-equib}
0&=-B_L\Gamma \sin (B^T \overline \theta)+p^\ast.
\end{align}
\ese 
Then, the point $\theta=\overline \theta$, $\omega_G=0$, and $u=\overline u$ identify an equilibrium of \eqref{e:sys-nl-ss}.
Let the right-hand side of \eqref{e:alg} be denoted by $g(\theta)$. To  investigate the regularity of \eqref{e:alg} and existence of the (local) solutions to the DAE \eqref{e:sys-nl-ss}, we compute the Jacobian of $g$ with respect to $\theta_L$ as
\beq\label{e:jacob}
\fpart{g}{\theta_L} =-B_L\Gamma [\cos(\eta)]B_L^T
\eeq
where $\eta=B^T \theta=B_G^T\theta_G+B_L^T\theta_L$, and $[\cos(\eta)]=\diag (\cos( \eta_k))$. 
Observe that the matrix $B_L\Gamma [\cos(\eta)]B_L^T$ is a principal submatrix of the Laplacian matrix
\[
L^\prime=B\Gamma^\prime(\eta) B^T 
\]
where 
\beq\label{e:gamma-prime}
\Gamma^\prime(\eta)=\Gamma [\cos(\eta)].
\eeq
Hence, $\fpart{g}{\theta_L}$ is nonsingular if $\Gamma'$ is positive definite. 
Therefore, the existence of an equilibrium and the regularity of \eqref{e:alg} is guaranteed under the following assumption:
\begin{assumption}\label{a:security}
\new{For given $\overline u$ and $p^\ast$}, there exists a constant vector $\overline \theta$ with $B^T \overline \theta \in \Omega:=(-\frac{\pi}{2}, \frac{\pi}{2})^m$ such that \eqref{e:ss-conditions} is satisfied.
\end{assumption}
The feasibility of the assumption above can be verified using the conditions proposed in \cite{dorfler2013synchronization}.
Under this assumption and considering a compatible initial condition, i.e.,
\beq\label{e:compat-theta}
0=-B_L\Gamma \sin (B^T\theta(0))+p^\ast
\eeq
the DAE \eqref{e:sys-nl-ss} admits a unique (local) solution, see \cite{Hill-DAE} for more details. Also note that the assumption $B^T \overline \theta \in \Omega$  is ubiquitous in the power grid literature and is sometimes referred to as a security constraint \cite{dorfler2014breaking}.

Next, we establish a reduced model for the system \eqref{e:sys-nl-ss}. 
Let $\eta=B^T\theta$ and $\omega=\col(\omega_G, \omega_L)$ as before. 
Then the differential algebraic system 
\eqref{e:sys-nl-ss} in the $\eta$ variables rewrites as
\bse\label{e:model-nl}
\begin{align}
\label{e:eta-nl}
\dot{\eta}&=B^T\omega=B_G^T\omega_G+B_L^T\omega_L\\
\label{e:w-nl}
M\dot\omega_G&=-A\omega_G-B_G\Gamma \sin (\eta)+u\\
\label{e:load-nl}
0&=-B_L\Gamma \sin (\eta)+p^\ast.
\end{align}
\ese
Note that solution $(\eta, \omega_G, \omega_L)$ of \eqref{e:model-nl} \new{with given $u\in\R^{n_g}$} are defined over the domain $\im B^T\times \R^{n_g}\times \R^{n_\ell}$. 
This set is obviously positive invariant, and coincides with the whole state space in case $\im B^T=\R^{n}$ meaning that the graph $\calG$ is acyclic. 
Taking the time derivative of \eqref{e:load-nl} along any solution $\eta$ yields
\begin{align}\label{e:time-d}
\nonumber
0&=-B_L\Gamma[\cos(\eta)]B^T\omega\\
&=-B_L\Gamma[\cos(\eta)]B_G^T\omega_G-B_L\Gamma[\cos(\eta)]B_L^T\omega_L
\end{align}
where $[\cos(\eta)]=\diag(\cos(\eta_k))$ as before.
Assuming that $\eta \in \Omega$, the matrix $[\cos(\eta)]$ is nonsingular, and  thus $\omega_L$ is obtained as
\beq\label{e:wl}
\omega_L=-(B_L\Gamma^\prime(\eta) B_L^T)^{-1}B_L\Gamma'(\eta) B_G^T\omega_G
\eeq
where $\Gamma^\prime$ is given by \eqref{e:gamma-prime}.
Note that by Assumption \ref{a:security} and equality \eqref{e:jacob} there exists a neighborhood around $\overline \eta=B^T\overline \theta$ such that $B_L\Gamma [\cos(\eta)]B_L^T$ is nonsingular, and there exists a solution to \eqref{e:sys-nl-ss}, and thus to \eqref{e:model-nl}, for a nonzero interval of time $\calI\subseteq \R_+$. This means that \eqref{e:time-d} and \eqref{e:wl} are well defined in this interval.

By substituting \eqref{e:wl} in \eqref{e:eta-nl}, we have
\beq\label{e:eta-new}
\dot\eta=(B_G^T-B_L^T(B_L\Gamma^\prime(\eta) B_L^T)^{-1}B_L\Gamma'(\eta) B_G^T)\omega_G.
\eeq
Now, let $B_S$ denote the projected incidence matrix with respect to the partitioning $B=\col(B_G, B_L)$ and the weight matrix $\Gamma^\prime(\eta)$ given by \eqref{e:gamma-prime}, i.e.
\[
B_S=B_G(I-B_L^+ B_L), \quad B_L^+(\eta)=\Gamma^\prime(\eta)(B_L\Gamma^\prime(\eta)B_L^T)^{-1}.
\]
Then, it is easy to see that the right-hand side of \eqref{e:eta-new} is equal to $B_S^T(\eta) \omega_G$, and hence we \new{obtain} the following reduced model
\bse\label{e:reduced}
\begin{align}
\label{e:eta-red}
\dot{\eta}&=B_S^T(\eta) \omega_G\\
\label{e:w-red}
M\dot\omega_G&=-A\omega_G-B_G\Gamma \sin (\eta)+u.
\end{align}
\ese
This defines a valid state space model for $(\eta, \omega_G)\in \im B^T \times \R^{n_g}$ with ordinary differential equations, and in particular we have the following theorem.
\begin{theorem}\label{t:relation}  
Considering the models \eqref{e:sys-nl-ss} and \eqref{e:reduced}, for given $u$ and $p^\ast$, the following statements hold:
\begin{enumerate}[(i)]
\item Let  $(\theta_G, \theta_L, \omega_G)$ be a solution to the differential algebraic equations \eqref{e:sys-nl-ss}, defined on the interval $\calI= [0, T)$. Assume that $B^T\theta \in \Omega$, $\theta=\col(\theta_G, \theta_L)$, $\forall t\in \calI$.
Then $(\eta, \omega_G)$ with $\eta=B^T\theta$ is a solution to the ordinary differential equations \eqref{e:reduced}, defined on the interval $\calI$.
\medskip{}
\item \new{Let $(\eta, \omega_G)$ be a solution to \eqref{e:reduced} on an interval $\calI=[0, T)$. 
Assume that $\eta(0)$ is a compatible initial condition for \eqref{e:alg}, i.e.
\beq\label{e:compat}
\eta(0)\in  \{v\in \im B^T \mid 0=p^\ast-B_L\Gamma \sin(v)\}.
\eeq
Then there exists a vector $\theta=\col(\theta_G, \theta_L)$ with $B^T\theta=\eta$ such that
 $(\theta_G, \theta_L, \omega_G)$ is a solution to \eqref{e:sys-nl-ss} on the interval $\calI$.}
\end{enumerate}
\end{theorem}
\BP 
The first statement of the theorem follows from the construction of the reduced model \eqref{e:reduced}. \new{The proof of the second statement requires an additional treatment and is deferred to a later moment.}
\EP

Theorem \ref{t:relation} promotes the system \eqref{e:reduced} as a legitimate reduced model for \eqref{e:sys-nl-ss}. By this theorem, the reduced network \eqref{e:reduced} and the original one {\eqref{e:sys-nl-ss} have identical behaviors once starting from the same and compatible initial conditions. 

 At the first glance, it seems that the constant power loads have disappeared in the reduced model \eqref{e:reduced}. However, these loads are actually embedded in the reduced dynamics. To see this, we make the following crucial observation, which is also relevant for completing the proof of Theorem \ref{t:relation}.
 
\begin{proposition}\label{p:conserved}
Let $\eta(0)\in \Omega$. 
Then the vector $B_L\Gamma \sin(\eta)$ is a conserved quantity of the dynamical system \eqref{e:reduced} over the domain $\calI$ of existence of the solution.
\end{proposition}
\BP
By taking the time derivative of $B_L\Gamma \sin(\eta)$ along the solutions of \eqref{e:reduced}, we obtain that
\begin{align*}
\frac{d}{dt}B_L\Gamma \sin(\eta)&=B_L\Gamma [\cos(\eta)]\dot{\eta}=
B_L\Gamma^\prime (\eta)B_S^T(\eta)\omega_G 
\end{align*}
Note that the matrix $\Gamma'$ is positive definite and the matrix $B_S$ is well defined in $\calI$. 
\new{By the second statement of Proposition  \ref{p:Bs} with $\Gamma$ being replaced by $\Gamma'(\eta)$, we find that $B_L\Gamma^\prime B_S^T(\eta)=0$ which completes the proof.}
\EP

Proposition  \ref{p:conserved} suggests that the constant vector $B_L\Gamma \sin(\eta)$ can indeed be interpreted as the constant power loads of the reduced network. \new{Notice that this vector has the same expression of the active power absorbed by the loads; see \eqref{e:active-i}}.

Assume that $u=\overline u$ is constant.
Then for an equilibrium $(\overline\eta, \overline\omega_G)$ of \eqref{e:reduced}, \new{necessarily} we have
\bse\label{e:equib}
\begin{align}
\label{e:equib-eta}
0&=B_S^T(\overline\eta) \overline\omega_G\\
\label{e:equib-w}
0&=-A\overline\omega_G-B_G\Gamma \sin (\overline\eta)+\overline u.
\end{align}
\ese
Hence,  by \eqref{e:equib-eta} and Proposition  \ref{p:Bs}(i),  we have $\overline\omega_G=\ones \omega^0$ for some constant $\omega^0$. By  multiplying both sides of \eqref{e:equib-w} from the left by $\ones^T$, we obtain that
$$
\omega^0=\frac{-\ones^TB_G\Gamma \sin (\overline\eta)+\ones^T\overline u}{\ones^TA\ones},
$$
which reduces to
\beq\label{e:omega0}
\omega^0=\frac{\ones^TB_L\Gamma \sin (\eta)+\ones^T\overline u}{\ones^TA\ones},
\eeq
where we have used the fact that $\ones^TB_G=-\ones^TB_L$, and $B_L\Gamma \sin (\eta)$ is constant. 
Hence, $\ones^T B_L\Gamma \sin (\eta)+\ones^T\overline u$ has to be identically zero to avoid frequency deviation. This corresponds to the well-known demand and supply matching condition which again elucidates the fact that the vector $B_L\Gamma \sin(\eta)$ plays the role of the loads in the reduced network \eqref{e:reduced}. 

By the discussion above, and the results of Theorem \ref{t:relation}(i) and Proposition \ref{p:conserved}, we conclude that the original network \eqref{e:sys-nl-ss} is {\em embedded} in the reduced network \eqref{e:reduced}. This enables us to deduce the properties of the original model by looking at the explicit reduced ODE model \eqref{e:reduced}.  We close this section by completing the proof of Theorem \ref{t:relation}, item $(ii)$.

\medskip{}

\noindent {\textbf{Proof of Theorem \ref{t:relation}}(ii) : 
First let $\omega_L$ be obtained from $\omega_G$ and $\eta$ using \eqref{e:wl}. Let the vector ${\delta}$ be such that $\eta(t)=B^T{\delta(t)}$.
To see such vector exists, note that by definition of $B_S$, we have $\ker B\subseteq \ker B_S$ and equivalently $\im B_S^T \subseteq \im B^T$. Hence, given the fact that $\eta(0)\in \im B^T$, we find that $\eta(t)\in \im B^T$ by \eqref{e:eta-red}.
We now define 
\beq\label{e:theta-proof}
\theta(t)={\delta}(t)- \ones \alpha(t)
\eeq
where $\alpha$ is given by
\beq\label{e:alpha-proof}
\alpha(t)=\frac{1}{n}(\ones^T\delta(t)-\ones^T \int_0^t \omega(\tau)d\tau)
\eeq
and $\omega=\col(\omega_G, \omega_L)$ as before. Notice that $B^T\theta=B^T\delta=\eta$ and hence $B^T\dot\theta=\dot{\eta}$.
In addition, from the derivation of \eqref{e:eta-red}, we have $\dot{\eta}=B^T\omega$. Therefore, $B^T\dot\theta=B^T\omega$ and thus
\[
\dot{\theta}=\omega+\ones \beta
\]
for some $\beta\in \R$. By \eqref{e:theta-proof}, the above writes as $\dot{\delta}-\ones \dot{\alpha}=\omega+\ones\beta$.
Hence, $\ones^T\dot{\delta}-n\dot{\alpha}=\ones^T\omega+n\beta$. By substituting \eqref{e:alpha-proof} in the latter equality, we conclude that $\beta=0$, and thus $\dot{\theta}=\omega$, and particularly \eqref{e:sys-nl-ss-thetaG} is satisfied. Clearly, \eqref{e:sys-nl-ss-omegaG} holds as well by $\eta=B^T\theta$, and it remains to show that the equality \eqref{e:alg} is satisfied.
Now by Proposition \ref{p:conserved}, the compatibility condition \eqref{e:compat} yields $B_L\Gamma \sin (\eta(t))=p^\ast$, for all 
$t\in \calI$. This in terms of $\theta$ variables writes as 
$B_L\Gamma \sin (B^T\theta(t))=p^\ast$, which completes the proof.
\EP
\vspace{-0.3cm}
\subsection*{Approximations of the reduced model:}
We recall that the reduced model \eqref{e:reduced} is not an {\em approximated} model of the power network \eqref{e:sys-nl-ss}, and in fact provides an alternative representation in terms of ordinary differential equations. However, if a simpler description of the network is needed, one can perform some approximations in \eqref{e:reduced}, and ultimately recover the linear reduced model \eqref{e:agg-new}. 
This will also highlight the relationship between the nonlinear reduced model \eqref{e:reduced} and the linear one \eqref{e:agg-new}.

The first approximation is to neglect the state-dependency in the dynamics of $\eta$ variables. Notice that this dependency is due to the matrix
\[
B_L^+(\eta)=\Gamma [\cos(\eta)] B_L^T (B_L \Gamma [\cos(\eta)] B_L^T)^{-1}.
\] 
Hence, in case the phase angles are almost uniform, the elements of $\eta$ are relatively small, and the matrix above can be approximated by the state-independent matrix $B_L^+$ in \eqref{e:BS-again}.
Consequently, \eqref{e:eta-red} will be replaced by 
\beq\label{e:etaS-new}
\dot{\eta}=B_S^T\omega_G.
\eeq
with $B_S$ given by \eqref{e:BS-again}. 
A second approximation is to replace $\sin(\eta)$ by $\eta$, which is again valid if the power network is working in a neighborhood of the nominal conditions and thus differences of the phase angles are relatively small. As a result of this, \eqref{e:w-red} rewrites as
\beq\label{e:approx-lin-omega}
M\dot\omega_G=-A\omega_G-B_G\Gamma \eta+u.
\eeq
Analogous to Proposition \ref{p:conserved}, it easy to see that the vector $B_L\Gamma \eta$ is a conserved quantity of \eqref{e:etaS-new}-\eqref{e:approx-lin-omega}, which we denote by the constant vector $p^\ast$:
\beq\label{e:nima-p*}
p^\ast=B_L\Gamma \eta(t).
\eeq  
Now the linear reduced model \eqref{e:agg-new} can be recovered from \eqref{e:etaS-new}-\eqref{e:approx-lin-omega} by a suitable projection:

\begin{proposition}
Suppose that $(\eta, \omega_G)$, with $\eta(0)\in \im B^T$, is a solution to \eqref{e:etaS-new}-\eqref{e:approx-lin-omega}. Define the vector $\eta_S=\Pi^T \eta$ where
\[
\Pi=I-B_L^+B_L=I-\Gamma B_L^T (B_L \Gamma B_L^T)^{-1}B_L,
\]
consistent with \eqref{e:Pi}. Then $(\eta_S, \omega_G)$ is a solution to \eqref{e:agg-new} where $\hatp$ is computed from \eqref{e:phat} with $p^\ast$ given by \eqref{e:nima-p*}.
\end{proposition} 
\BP
By definition of $B_S$, it follows that
\[
\ker \bbm B_S \\B_L \ebm =\ker B.
\]
Now, as $\eta(0)\in \im B^T$, similar to the proof of Theorem \ref{t:relation}(ii) we find that $\eta(t)\in \im B^T$. Hence, the vector $\eta$ can be written as 
\beq\label{e:eta-decompose}
\eta=B_S^Tz_S+B_L^Tz_L
\eeq
for some vectors $z_S$ and $z_L$.
By multiplying both sides of the latter equality with the matrix $B_L\Gamma$, and using \eqref{e:nima-p*} together with the second item of Proposition \ref{p:Bs} we find that
\[
z_L=(B_L\Gamma B_L)^{-1} p^\ast.
\]
Substituting \eqref{e:eta-decompose}, with $z_L$ given above, into \eqref{e:approx-lin-omega} yields
\[
M\dot\omega_G=-A\omega_G-B_G\Gamma B_S^Tz_S+u- \hat{p}
\]
where $\hat{p}$ has the same expression as in \eqref{e:phat}.
Now, the vector $B_S^Tz_S$ is computed as  
\begin{align*}
B_S^Tz_S&=\eta-B_L^Tz_L=\eta-B_L^T(B_L\Gamma B_L)^{-1} p^\ast\\
&=\eta-B_L^T(B_L\Gamma B_L)^{-1} B_L\Gamma \eta=\Pi^T\eta=\eta_S.
\end{align*}
Consequently,  $(\eta_S, \omega_G)$ is a solution to
\begin{align*}
\dot{\eta}_S&=B_S^T\omega_G\\
M\dot{\omega}_G&=-A\omega_G-B_G \Gamma \eta_S+u-\hatp.
\end{align*}
The system above is identical to the linear reduced model \eqref{e:agg-new} by exploiting the identities
\[
B_G \Gamma \eta_S=B_G \Gamma B_S^Tz_S=B_S \Gamma B_S^Tz_S=B_S \Gamma\eta_S
\]
where we used Proposition \ref{p:Bs} to write the second equality.
\EP}

\section{Analysis and Control of the reduced model}\label{s:convergence}

In this section, we show the practicality and usefulness of the established reduced model for analysis and design purposes including frequency regulation and active power sharing. Although the reduced network \eqref{e:reduced} is expressed in terms of ordinary differential equations,
 the existing control schemes are not readily applicable to this case \cite{trip2016internal, Claudio-Nima-modular, monshizadeh2015output}. 
In particular, the map $B_S$ in \eqref{e:eta-red} is state-dependent unlike the ordinary time-independent incidence matrix. 
In addition, different to the linear model \eqref{e:ph-structure},  it is easy to see that the underlying Poisson structure of \eqref{e:reduced} is not defined on a skew-symmetric matrix, and thus the stability/passivity of the system does not readily follows from standard port-Hamiltonian reformulation of the system \cite{van2014port,schaft2013}. 
However, one can show that this does not hinder the analysis thanks to the remarkable properties of the projected incidence matrix captured in Proposition  \ref{p:Bs} as well as the invariance property highlighted in Proposition  \ref{p:conserved}.  This will be elaborated in the current section. 


\new{To conclude stability properties of \eqref{e:reduced} and pave the way for a controller design at the same time, we first establish the passivity property of an incremental representation of \eqref{e:reduced}.} Recall the equalities \eqref{e:equib} with $\overline \eta\in \Omega \cap \im B^T$, $\overline \omega_G=\ones \omega^0$, and $\omega^0$ given by \eqref{e:omega0}. By Proposition \ref{p:conserved} and given $u=\overline u$, the pair $(\overline \eta, \overline \omega_G)$ is a valid steady-state solution of the system only if 
\beq\label{e:compat-nima}
B_L\Gamma \sin (\eta(0))=B_L\Gamma \sin (\overline \eta).
\eeq
This is the same compatibility condition assumed in \eqref{e:compat-theta}, noting \eqref{e:alg-equib}. 
Existence of such $\overline \eta$ and $\overline \omega_G$ is accounted for the feasibility condition. In fact, we shall assume that there exists $\overline \eta\in \Omega \cap \im B^T$ such that \eqref{e:equib-w} and \eqref{e:compat-nima} are satisfied, with $\overline \omega_G=\ones \omega^0$. This condition is similar to the one in Assumption \ref{a:security} and can be verified for instance using the result of \cite{dorfler2013synchronization}.
Now we write the dynamics \eqref{e:reduced} as
\bse\label{e:incremental}
\begin{align}
\label{e:incremental-eta}
\dot{\eta}&=B_S^T(\eta)(\omega_G-\overline \omega_G)\\
\label{e:incremental-omega}
M\dot\omega_G&=-A(\omega_G-\overline \omega_G)-B_G\Gamma (\sin (\eta)-\sin (\overline\eta))+u-\overline u.
\end{align}
\ese
where in \eqref{e:incremental-eta} we have used the fact that $\overline \omega_G\in \im \ones$ and thus $B_S^T(\eta)\overline \omega_G$ is equal to zero by Proposition \ref{p:Bs}(i). The equality \eqref{e:incremental-omega} is written by exploiting \eqref{e:equib-w}.
Now, similar to \cite{trip2016internal,monshizadeh2015output,monshizadeh2016agreeing},  let the storage function $W$ be defined as
\begin{align}\label{e:W}
\nonumber
&W(\eta, \omega_G)=\frac{1}{2} (\omega_G-\overline\omega_G)^TM(\omega_G-\overline\omega_G)\\
&\quad +\ones ^T\Gamma \cos (\eta)-\ones ^T\Gamma \cos (\overline\eta)-(\Gamma \sin(\overline\eta))^T(\eta-\overline\eta)
\end{align}
First, notice that $W(\overline \eta, \overline \omega_G)=0$. In addition, $(\overline \eta, \overline \omega_G)$ constitutes a strict minimum of
$W$ in  $\Omega\times \R^{n_g}$. In particular, the partial derivatives of $W$ are computed as
\beq\label{e:part-eta}
\fpart{W}{\eta}=\Gamma (\sin (\eta)-\sin (\overline \eta))
\eeq
and
\beq\label{e:part-omega}
\fpart{W}{\omega_G}=M(\omega_G-\overline \omega_G)
\eeq
which vanish at $( \eta, \omega_G)=(\overline \eta, \overline \omega_G)$.
The Hessian of $W$ is given by the matrix 
\[
\bdiag\;(\Gamma [\cos(\eta)], M),
\]
which is positive definite in $\Omega\times \R^{n_g}$.
Now, passivity of \eqref{e:incremental} with output variables $y=\omega_G-\overline \omega_G$ follows from the following proposition:
\begin{proposition}\label{p:passive}
Let $W$ be defined as in \eqref{e:W}. Then the time derivative of $W$ along the solution $(\eta, \omega_G)$ of \eqref{e:reduced}, initialized in a neighborhood of $(\overline \eta, \overline \omega_G)$ with $B_L\Gamma \sin (\eta(0))=B_L\Gamma \sin (\overline \eta)$, satisfies the following dissipation equality
\begin{align}\label{e:passivity}
\nonumber
\dot{W}=&-(\omega_G-\overline\omega_G)^TA(\omega_G-\overline\omega_G)\\
&\quad +(\omega_G-\overline\omega_G)^T(u-\overline u)
\end{align}
for the interval of definition $\calI$ of the solution.
\end{proposition}
\BP
Recall that \eqref{e:reduced} can be equivalently written as \eqref{e:incremental}. By using \eqref{e:part-eta} and \eqref{e:part-omega}, we obtain that
\begin{align*}
\dot{W}&=(\sin (\eta)-\sin (\overline \eta))^T\Gamma B_S^T(\eta)(\omega_G-\overline \omega_G)\\&-(\omega_G-\overline \omega_G)^TA(\omega_G-\overline \omega_G)
+(\omega_G-\overline\omega_G)^T(u-\overline u)\\
&-(\omega_G-\overline \omega_G)^TB_G\Gamma (\sin (\eta)-\sin (\overline \eta))
\end{align*}
Hence, it suffices to show that 
\beq\label{e:claim-passivity}
(\omega_G-\overline \omega_G)^T(B_S(\eta)-B_G)\Gamma  (\sin (\eta)-\sin (\overline \eta))=0.
\eeq
for all $t\in\calI$.
Recall that
\[
B_S(\eta)=B_G-B_G\Gamma^\prime B_L^T(B_L\Gamma^\prime B_L^T)^{-1}B_L
\]
with $\Gamma^\prime=\Gamma [\cos(\eta)]$. Then \eqref{e:claim-passivity} holds if 
\beq\label{e:claim-passivity-2}
B_L\Gamma  (\sin (\eta)-\sin (\overline \eta))=0.
\eeq
As $B_L\Gamma \sin (\eta(0))=B_L\Gamma \sin (\overline \eta)$, the above reduces to $B_L\Gamma  \sin (\eta)=B_L\Gamma\sin (\eta(0))$ which holds true by Proposition \ref{p:conserved}. This completes the proof. \EP

Now, by using Proposition \ref{p:passive}, attractivity of the equilibrium $(\overline \eta, \overline\omega_G)$ is established next. 

\begin{theorem}\label{t:attractivity}
Let $u=\overline u$ with a constant vector $\overline u$, and assume that $(\overline\eta, \overline \omega_G)$ with $\overline \eta\in \Omega \cap \im B^T$ is an equilibrium of \eqref{e:reduced}. 
Then solutions $(\eta, \omega_G)$ of \eqref{e:reduced} with $B_L\Gamma \sin (\eta(0))=B_L\Gamma \sin (\overline \eta)$ locally converge to $(\overline \eta, \overline \omega_G)$.  
\end{theorem}
\BP
By  \eqref{e:passivity}, we obtain $\dot{W}=-(\omega_G-\overline\omega_G)^TA(\omega_G-\overline\omega_G)$.
Noting again that $(\overline \eta, \overline \omega_G)$ is a (local) strict minimum of $\calW$, one can construct a compact level set around 
$(\overline \eta, \overline \omega_G)\in (\im B^T\cap \Omega)\times \R^{n_g}$ which is forward invariant. By invoking LaSalle's invariance principle on the invariant set we have 
$\omega_G=\overline \omega_G$ and
\bse\label{e:invariant-set}
\begin{align}
\dot{\eta}&=B_S^T(\eta) \omega_G=B_S^T(\eta) \overline\omega_G=0\\
\label{e:w-inv}
0&=-A\overline\omega_G-B_G\Gamma \sin (\eta)+\overline u.
\end{align}
\ese
By \eqref{e:equib-w}, the equality \eqref{e:w-inv} yields $
B_G\Gamma \sin (\eta)=B_G\Gamma \sin(\overline \eta).$
Moreover, again by Proposition \ref{p:conserved}, and the fact that $B_L\Gamma \sin (\eta(0))=B_L\Gamma \sin (\overline \eta)$, the equality \eqref{e:claim-passivity-2} holds. Therefore, by continuity
\beq\label{e:inv-sin-eta}
B\Gamma (\sin (\eta)-\sin(\overline \eta))=0
\eeq
on the invariant set. Now, as $\eta, \overline \eta \in \im B^T$, we have $\eta=B^T\theta$ and $\overline \eta=B^T \overline \theta$ for some vectors $\theta, \overline \theta \in \R^n$. Then multiplying \eqref{e:inv-sin-eta} from the left by $(\theta-\overline \theta)^T$ gives $(\eta-\overline \eta)^T\Gamma (\sin (\eta)-\sin(\overline \eta))$. Since $\sin(\cdot)$ is strictly monotone in $\Omega$, we conclude that $\eta=\overline \eta$ on the invariant set, which completes the proof.
\EP

Next, we include a controller in order to regulate the frequency deviation to zero while achieving certain power sharing properties.  
By Theorem \ref{t:attractivity}, for $u=\overline u$, the solutions of \eqref{e:reduced} locally converge to a common steady-state frequency identified by $\overline \omega_G=\ones \omega^0$, where $\omega^0$ is calculated as in \eqref{e:omega0}. 
A nonzero $\omega^0$ indicates a static deviation from nominal frequency, which must be eliminated.
The choice of $\overline u$ to eliminate this deviation is in general not unique. The corresponding degree of freedom can be leveraged to achieve an optimal deployment of the control effort. In particular, similar to \cite{trip2016internal, dorfler2014breaking}, we consider the following \emph{optimal frequency regulation problem}: \begin{subequations}\label{e:OFR}
\begin{align}
\label{e:cost}
& \underset{\overline u \in \R^{|\calV_G|}}{\text{minimize}}
& & \frac{1}{2}\overline u^TQ\overline u=\sum_{i\in\mathcal{V}_G}\nolimits \frac{1}{2}q_i\overline u_i^2 \\
\label{e:match}
& \text{subject to}
& 0&=\ones^T \overline u +\ones^Tp^\ast,
\end{align}
\end{subequations}
where \new{$p^\ast:=B_L\Gamma \sin (\overline \eta)$}, and we minimize the total quadratic cost of generation \eqref{e:cost} subject to the power balance \eqref{e:match}. 
Here, $Q=\diag(q_i)$ with $q_i\in\R_{+}$ being the cost coefficient, and $\frac{1}{2} q_i u_{i}^2$ being the local generation cost at the $i${th} generator.  Notice that \eqref{e:match} amounts to matching ``supply" and ``demand", and enforces the zero frequency deviation. 
Following the standard Lagrange multipliers method, the optimal control $u^\star_{i}$ that minimizes \eqref{e:cost} subject to the constraint \eqref{e:match} is computed as
\beq\label{e:ui-optimal}
u^\star_i =-\lambda q^{-1}_i
\eeq
where
\begin{equation}
\nonumber
\lambda=\frac{\ones^T p^\ast}{\sum_{i\in\calV_G} q_i^{-1}}
\end{equation}
is the multiplier of the constraint \eqref{e:match},  and can be interpreted as the ``price" per unit of generation. The equality \eqref{e:ui-optimal} implies that $u_i^\star q_i=u_j^\star q_j$ for all $i,j \in \calV_G$, meaning that the generators should provide power at {\em identical marginal costs}.

Note that substituting the zero frequency deviation $\overline \omega_G=0$ and optimal control $\overline u=u^\star:=\col(u^\star_i)$ in \eqref{e:equib} yields  
\beq\label{e:new-assump}
0=-B_G\Gamma \sin (\overline\eta)-\lambda Q^{-1}\ones.
\eeq
Similar to before, we assume that the above equality has a solution $\overline \eta \in \im B^T \cap \Omega$. This is the same as Assumption \ref{a:security} by setting $\overline u=u^{\star}$.

To avoid centralized information in \eqref{e:ui-optimal}, and to distribute the solution of the optimal control deployment problem in real-time, distributed averaging integral controllers have been proposed in the literature \cite{JWSP-FD-FB:12u, dorfler2014breaking, trip2016internal}; see also {\cite{AJ-ML-PPJVDB:09,QS-JG-JMV:13,MA-DVD-KHJ-HS:13,AB-FLL-AD:14} for related work on distributed secondary frequency controllers.} These controllers are defined on a connected communication graph $\calG_c=(\calV_c, \calE_c)$ and take the form
\bse\label{e:controller-i}
\begin{align}
\dot{\xi}_i&=-\sum_{\{i,j\}\in\mathcal{E}_{\rm c}} (\xi_i-\xi_j)-q_i^{-1}\omega_i\\
 u_i&=q_i^{-1}\xi_i
\end{align}
\ese
for each $i\in \calV_G$. Here, the state $\xi_i$ acts as a local copy of the multiplier $\lambda$ for each unit, the term $q_i^{-1}\omega_i$ attempts to regulate the frequency deviation to zero, and the consensus based algorithm $\sum_{\{i,j\}\in \mathcal{E}_{\rm c}} (\xi_i-\xi_j)$ enforces identical marginal costs at steady-state.

The controller above can be written in the vector form as
\bse\label{e:controller}
\begin{align}
\dot{\xi}&=-L_ C\xi-Q^{-1}\omega_G\\
 u&=Q^{-1}\xi
\end{align}
\ese
where $L_C$ is the Laplacian matrix of $\calG_c$, and $\xi=\col(\xi_i)$, $Q=\diag(Q_i)$, with $i\in \calV_G$. 
It is easy to see that $\xi=Qu^\star$ and $\omega_G=0$ is a solution to \eqref{e:controller}. 
Interconnecting this controller to the model \eqref{e:reduced} results in a zero frequency deviation and optimal deployment of the active power 
\eqref{e:ui-optimal} as shown in the following theorem:
\begin{theorem}
Let $\overline \omega_G=0$, $\overline \xi=Qu^\star$, and assume that the vector $\overline \eta$ is such that \eqref{e:new-assump} holds.
Consider the model \eqref{e:reduced} in closed loop with \eqref{e:controller}. 
Then solutions $(\eta, \omega_G, \xi)$, with $B_L\Gamma \sin(\eta(0))=B_L\Gamma \sin (\overline \eta)$, 
locally converge to
the point $(\overline \eta, \overline \omega_G, \overline \xi)$.
Consequently, the vector $u$ locally converges to the optimal input $u^\star$ given by \eqref{e:ui-optimal}.
\end{theorem}

\BP
Let $W_C$ be defined as $W_C(\xi)=\frac{1}{2}(\xi-\overline \xi)^T(\xi-\overline \xi)$. Then the time derivative of $W_C$ along the solutions of \eqref{e:controller} is computed as
\begin{align*}
\dot{W}_C&=-(\xi-\overline \xi)^TL_C\xi-(\xi-\overline \xi)^TQ^{-1}\omega_G\\
&=-(\xi-\overline \xi)^TL_C(\xi-\overline \xi)-(u-u^\star)^T(\omega_G-\overline \omega_G)
\end{align*}
where we have used the facts that $\overline \xi=-\lambda \ones$ and $\overline \omega_G=0$ to write the second equality.
Now, let
\[
V(\eta, \omega_G, \xi)=W(\eta, \omega_G)+W_C(\xi).
\]
Then, by \eqref{e:passivity} and noting $\overline u=Q^{-1}\overline \xi=u^\star$, the time derivative of $V$ along the solutions of \eqref{e:reduced},\eqref{e:controller}, is calculated as
\[
\dot{V}=-(\omega_G-\overline\omega_G)^TA(\omega_G-\overline\omega_G)-(\xi-\overline \xi)^TL_C(\xi-\overline \xi).
\]
Noticing that $(\overline \eta, \overline \omega, \overline \xi)$ is a (local) strict minimum of $V$, by applying LaSalle's invariance principle we obtain $\omega_G=\overline \omega_G=0$, and
\begin{align*}
0&=L_C(\xi-\overline \xi)\\
0&=-B_G\Gamma \sin (\eta)+Q^{-1}\xi.
\end{align*}
on the invariant set. The first equality gives $\xi=\overline \xi+\alpha \ones$ for some $\alpha \in \R$. By multiplying both sides of the second equality with $\ones^T$, we obtain 
\beq\label{e:proof-controller-inv}
0=\ones^TB_L\Gamma \sin (\eta)+\ones^TQ^{-1}\overline \xi+\alpha \ones^TQ^{-1}\ones.
\eeq
By Proposition \ref{p:conserved} and continuity, $B_L\Gamma \sin (\eta(0))=B_L\Gamma \sin (\eta)$. Hence, the equality \eqref{e:proof-controller-inv} can be rewritten as
\[
0=\ones^TB_L\Gamma \sin (\overline \eta)+\ones^Tu^\star+\alpha \ones^TQ^{-1}\ones.
\]
Noting that $\overline u=u^\star$ given by \eqref{e:ui-optimal} satisfies \eqref{e:match}, we conclude that $\alpha=0$, and thus $\xi=\overline \xi$ and $u=u^\star$ on the invariant set. Analogous to the proof of Theorem \ref{t:attractivity}, $\eta$ also coincides with $\overline \eta$ on this invariant set, and the proof is complete.
\EP

We close this section by a numerical example.
\vspace{-0.2cm}
\subsection{Case study}
We consider the power network model \cite{wood1996power} consisting of three generators and three loads as shown in Figure \ref{f:topology}. 
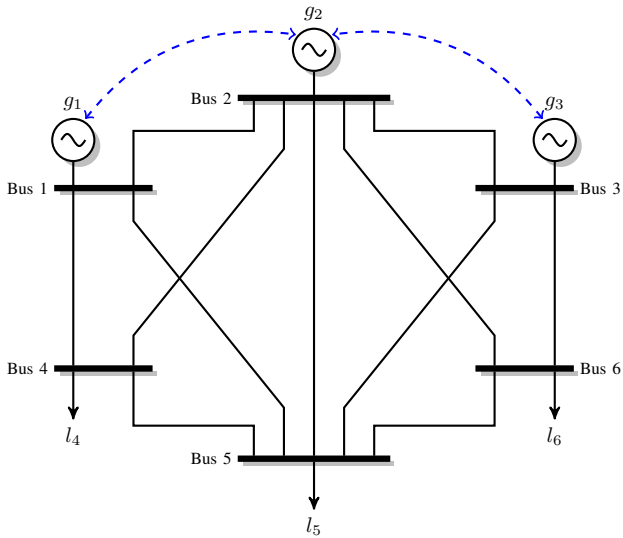
\begin{figure}\label{f:topology}
\medskip
\centering
\begin{tikzpicture}[thick, scale=0.8, transform shape]

\tikzstyle{line}=[-, thick]
\tikzstyle{loadline}=[->,thick,>=stealth']
\tikzstyle{busbar} = [rectangle,draw,fill=black,inner sep=0pt];
\tikzstyle{hbus} = [busbar,minimum width=10mm,minimum height=2pt,drop shadow];

\coordinate (c1) at (0,3);
\coordinate (c2) at (3.5,4.5);
\coordinate (c3) at (7,3);
\coordinate (c4) at (0,0);
\coordinate (c5) at (3.5,-1.5);
\coordinate (c6) at (7,0);
\coordinate (over) at (0,3.75);

\begin{footnotesize}
\node[hbus,minimum width=16mm,label=left:Bus 1] (b1) at (c1) {};
\node[hbus,minimum width=25mm,label=left:Bus 2] (b2) at (c2) {};
\node[hbus,minimum width=16mm,label=right:Bus 3] (b3) at (c3) {};
\node[hbus,minimum width=16mm,label=left:Bus 4] (b4) at (c4) {};
\node[hbus,minimum width=25mm,label=left:Bus 5] (b5) at (c5) {};
\node[hbus,minimum width=16mm,label=right:Bus 6] (b6) at (c6) {};
\end{footnotesize}

\draw[line,<->, dashed, blue] ([xshift=-3mm, yshift=11mm] b1.north) to[bend left] ([xshift=-3mm, yshift=11mm] b2.south);
\draw[line,<->, dashed, blue] ([xshift=3mm, yshift=11mm] b3.north) to[bend right] ([xshift=3mm, yshift=11mm] b2.south);
\draw[line] ([xshift=5mm] b1.north) |- ([xshift=-10mm,yshift=-5mm] b2.south)  --
([xshift=-10mm] b2.south);
\draw[line] ([xshift=-5mm] b1.south) -- ([xshift=-5mm] b4.north) ;
\draw[line] ([xshift=5mm] b1.south) -- ([xshift=5mm,yshift=-5mm] b1.south)
-- ([xshift=-5mm,yshift=8mm] b5.north) -- ([xshift=-5mm] b5.north);
\draw[line] ([xshift=10mm] b2.south) -- ++(0,-0.5) -| ([xshift=-5mm]
b3.north);
\draw[line] ([xshift=-5mm] b2.south) --
([xshift=-5mm,yshift=-8mm] b2.south) -- ([xshift=5mm,yshift=5mm] b4.north)
-- ([xshift=5mm] b4.north);
\draw[line] (b2.south) -- (b5.north);
\draw[line] ([xshift=5mm] b2.south) -- ([xshift=5mm,yshift=-8mm] b2.south)
-- ([xshift=-5mm,yshift=5mm] b6.north) -- ([xshift=-5mm] b6.north);
\draw[line] ([xshift=-5mm] b3.south) -- ([xshift=-5mm,yshift=-5mm] b3.south)
-- ([xshift=5mm,yshift=8mm] b5.north) -- ([xshift=5mm] b5.north);
\draw[line] ([xshift=5mm] b3.south) -- ([xshift=5mm] b6.north);
\draw[line] ([xshift=5mm] b4.south) |- ([xshift=-10mm,yshift=5mm] b5.north) --
([xshift=-10mm] b5.north);
\draw[line] ([xshift=10mm] b5.north) -- ([xshift=10mm,yshift=5mm] b5.north) -|
([xshift=-5mm] b6.south);

\genset{g1}{$(c1)+(-5mm,8mm)$}{label=above:$g_1$}
\draw[line] ([xshift=-5mm] b1.north) -- (g1.south);
\genset{g2}{$(c2)+(0,8mm)$}{label=above:$g_2$}
\draw[line] (b2.north) -- (g2.south);
\genset{g3}{$(c3)+(5mm,8mm)$}{label=above:$g_3$}
\draw[line] ([xshift=5mm] b3.north) -- (g3.south);

\draw[loadline] ([xshift=-5mm] b4.south) -- ++(0,-0.8) node[text centered,text
width=10mm,below] {$l_4$};
\draw[loadline] (b5.south) -- ++(0,-0.8) node[text centered,text
width=10mm,below] {$l_5$};
\draw[loadline] ([xshift=5mm] b6.south) -- ++(0,-0.8) node[text centered,text
width=10mm,below] {$l_6$};

\end{tikzpicture}
\caption{Diagram for a 6-bus power network, consisting of 3 generator and 3 load buses. The communication links of the secondary controller \eqref{e:controller-i} are represented by the dashed lines.}
\label{fig:case6ww}
\end{figure}
\begin{figure}[t!]
\centering
\includegraphics[width = .49\textwidth]{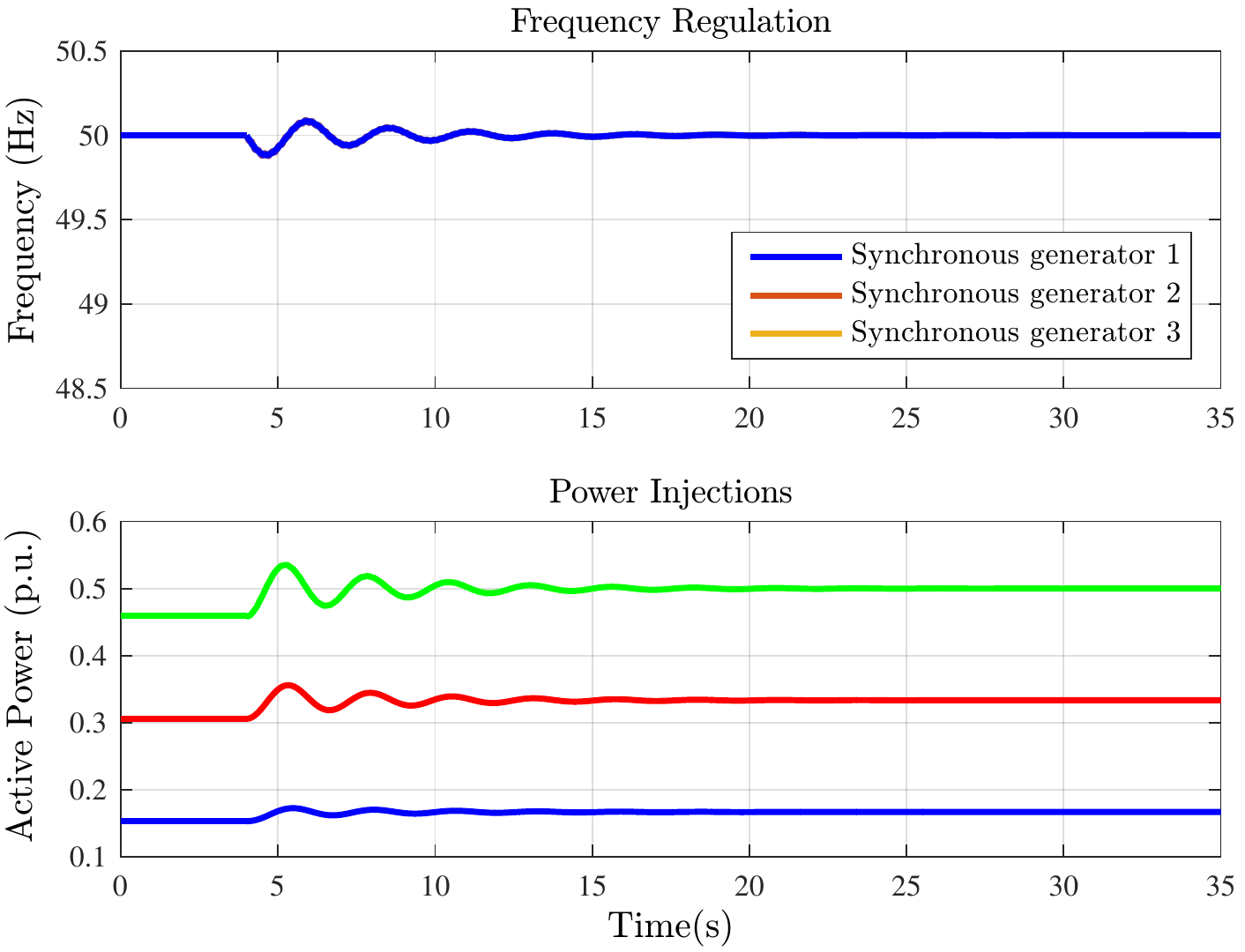}
\caption{Numerical simulation of the results}\label{f:sim}
\end{figure}
The power network parameters are chosen as: $M_{1}=4.62$, $M_{2}=4.17$, $M_3=5.10$, $A_{1}=1.41$, $A_2=1.28$, $A_3=1.72$. The line inductances (pu) are given by 
\begin{align*}
&{\small{[X_{12},X_{14},X_{15},X_{23},,X_{24},X_{25},X_{26},X_{35},X_{36},X_{45},X_{56}]}=}\\
&[0.25,0.21,0.32,0.26,0.13,0.33,0.22,0.31,0.10,0.50,0.33].
\end{align*}

The quadratic cost function in \eqref{e:cost} is considered as $Q=0.4\diag(1, \frac{1}{2}, \frac{1}{3})$.  We employ the controllers \eqref{e:controller} at the synchronous generators. The system is initially at steady-state with constant power loads. At time $t=4 \;\rm sec$, the active power load at Bus $4$ is increased by 20 percent of its nominal value. As we consider a constant power load model, this increase results in  a step in the phase angle of Bus $4$, and thus affects the frequency response of the system \eqref{e:reduced} as can be seen from Figure \ref{f:sim}. The frequency evolution and the active power injections of \eqref{e:reduced} in closed-loop with \eqref{e:controller} are depicted in Figure \ref{f:sim}. It is observed that the controller restores the frequency at $50\; \rm Hz$ (the frequencies at the various buses are so similar to each other that no difference can be noticed in the plot). In addition, the generation costs are minimized meaning that power is proportionally shared according to the cost coefficients given by the diagonal elements of $Q$ (with the ratio of $1$, $2$, and $3$), consistent with \eqref{e:ui-optimal}.

\section{Conclusions}\label{s:concl}
We have considered structure preserving power networks expressed as differential algebraic equations, where
the proper algebraic constraints are the result of the presence of constant power loads.
We have introduced the notion of the projected incidence matrix, which provides a novel decomposition of the reduced Laplacian matrix. For the linear network model, by exploiting this new matrix, we have derived a novel reduced model which preserves many structural properties of the original network. We have also addressed the elimination of algebraic constraints in the nonlinear network model.
Again, by using the projected incidence matrix, we have established a reduced model under a suitable regularity assumption. The reduced model is expressed in terms of ordinary differential equations, and thus facilitates the analysis, controller design, and simulation of the power network. Frequency regulation and active power sharing of the reduced model are addressed by using a distributed averaging controller. Extension of the proposed reduction technique to time-varying voltages and coupled power flow will be studied in future. Other future research directions include investigating the use of the projected incidence matrix in other model reduction techniques which are based on Schur complements of the Laplacian matrix, see. e.g. \cite{van2016network}. Possible relation and applications to clustering \cite{besselinkclustering,ishizaki2014model,ishizaki2016clustered,Nima-clust,monshizadeh2014structure}, and slow coherency, see e.g. \cite{romeres2013novel,biyik2008area,chow1985time} also deserve attention.
\bibliographystyle{IEEEtran}
\bibliography{ref,nima2,MainRefs,FD,JWSP,Main,alias,New}
\end{document}